\documentclass[aps,twocolumn,prb,showpacs,balancelastpage,amssymb,groupedaddress]{revtex4}

\usepackage{graphicx}%
\usepackage{epsfig}

\begin{document}

\title{ Current rectification {\bf by} asymmetric molecules: \centerline {An {\sl ab initio} study} }%

\author{Yan-hong Zhou$^{1}$}

\author{Xiao-hong Zheng $^2$}

\author{ Ying Xu$^{1}$}
\email {yingxu@jxnu.edu.cn}

\author{Zhao Yang Zeng$^{1}$}

\affiliation { $^1$Department of Physics, Jangxi Normal University, Nanchang 330022, China\\
$^2$Key Laboratory of Material Physics, Institute of Solid State
Physics, Chinese Academy of Sciences, Hefei, Anhui 230031, China}

\date{\today}

\begin{abstract}

We study current rectification effect in an asymmetric molecule
HOOC-C$_6$H$_4$-(CH$_2$)$_n$ sandwiched between two Aluminum
electrodes using an {\sl ab initio} nonequilibrium Green function
method. The conductance of the system decreases exponentially with
the increasing number $n$ of CH$_2$. The phenomenon of current
rectification is observed such that a very small current appears
at negative bias and a sharp negative differential resistance at a
critical positive bias when $n\ge 2$. The rectification effect
arises from the asymmetric structure of the molecule and the
molecule-electrode couplings. A significant rectification ratio of
$\sim$38 can be achieved when $n=5$.

\end{abstract}
\pacs{85.65.+h, 73.63. -b, 36.40.-c \\[-5mm]}

\maketitle

\section{INTRODUCTION}

Electronic transport properties of single molecule junctions have
gained tremendous interest in recent years, since they may have a
wide variety of important applications in future electronic
components such as transistors, diodes and switches
\cite{1,2,3,4,5}. With the advantages of experimental techniques,
for example, scanning tunneling microscope and mechanically
controllable break junction\cite{6,7,8},  measurement of the
current through nanoscale systems is now allowed. Some interesting
behaviors, such as highly nonlinear $I$-$V$ characteristics,
negative differential resistance (NDR) and electric switching
behavior, are observed in various systems \cite{9,10,11}.
Meanwhile, considerable amounts of theoretical work have been
performed to study transport properties of molecular devices.

\begin{figure}
\epsfig{figure=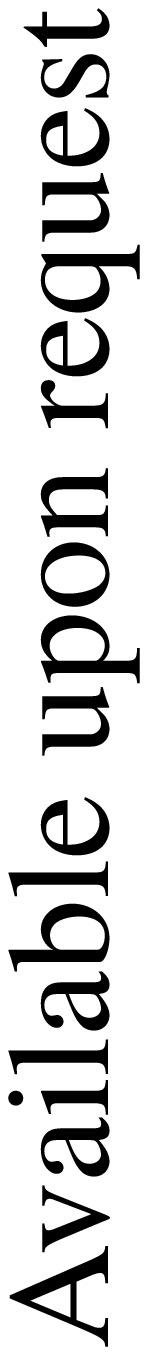,width=3.0in}
 \caption{(color online) Model structure of a two-probe
system with a single molecule HOOC-C$_6$H$_4$-(CH$_2$)$_n$-S
coupled to two Al(100) electrodes.}
\end{figure}

\begin{figure}
\epsfig{figure=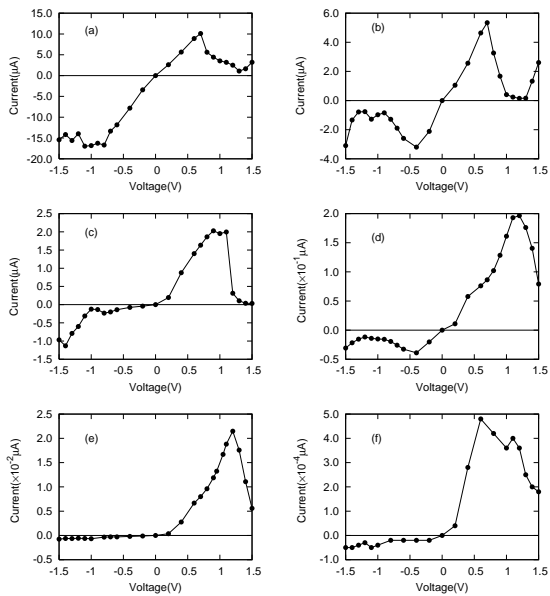,width=3.0in} \caption{ The (I-V) curves for
different number of CH$_2$: (a) $n=0$, (b) $n=1$, (c) $n=2$, (d)
$n=3$, (e) $n=5$ and (f) $n=7$. }
\end{figure}

Rectification, one of the most important functions in a
traditional electronic component, has also been suggested and
observed in molecular devices. Some mechanisms for rectification
phenomena in these devices have been suggested\cite{12,13,14,15}.
The first molecular rectifier was proposed by Aviram and Ratner in
1974 using $D$-$\sigma$-$A$ molecules, where $D$ and $A$ are,
respectively, an electron donor and an electron acceptor, and
$\sigma$ is a covalent ``sigma'' bridge(insulator). In this model,
the inelastic electron transfer is more favorable from $A$ to $D$,
rather than in the opposite direction, thus current rectification
happens. However, electrical rectification was observed
experimentally only recently in Langmuir-Blodgett (LB) multilayers
or monolayers of the molecule C$_{16}$H$_{33}$Q-3CNQ sandwiched
between metal electrodes \cite{16,17,18}. The underlying
mechanism, which was first considered to be a possible
implementation of the Aviram-Ratner mechanism, is of somewhat
different $D$-$\pi$-$A$ type. Although the molecule does show
rectification, it behaves as an anisotropic insulator rather than
a conductor because of the long alkane tail C$_{16}$H$_{33}$. In
fact, the current is of order  10$^{-17}A/molecule$ for the
structure Al/C$_{16}$H$_{33}$Q-3CNQ/Al and $10^{-15}$A/molecule
for the structure Au/ C$_{16}$H$_{33}$Q-3CNQ/Au in Ref.\cite{9}.
The current is too small for practical applications. To achieve
effective rectification with a reasonably conductive molecule, one
should avoid using molecules with long saturated groups that may
lead the molecules to be prohibitively resistive. It seems
attractive to use relatively short molecules with certain end
groups to allow their selfassembly on a metallic electrode's
surface. In Ref. $19$, transport measurement on the structure of
two phenyl rings connected to electrodes with asymmetric alkane
chains (molecule (CH$_2$)$_m$-C$_{10}$H$_6$-(CH$_2$)$_n$) has been
performed\cite{19}. It suggests another new mechanism of molecular
rectification, where a single electroactive unit is positioned
asymmetrically with respect to electrodes and the highest occupied
molecular orbital(HOMO) and lowest unoccupied molecular
orbital(LUMO) are positioned asymmetrically with respect to the
Fermi level. For majority of the applied voltage drops on the
longer insulating barrier, the conditions for resonant tunneling
through the active level are different for different directions of
the applied voltages on the two opposite polarities. A maximal
rectification ratio $35$ is expected with longer alkane chains
$n=10$ and $m=2$, the current magnitude is still not ideal, with
order $10^{-10}A$. Recently ideal rectification has been
demonstrated in diblock oligomer diode molecules \cite{14} and
single C$_{59}$N molecules \cite{20}. However, the current
reported\cite{14,17} is still of the unsatisfactorily order
$10^{-10}A$. In this work, we perform a systematic study on the
transport properties of a single molecule
HOOC-C$_6$H$_4$-(CH$_2$)$_n$. Some of our findings are
interesting. First, a significant rectification ratio of $\sim$38
is achieved at bias 1.2$V$ for much shorter alkane tail ($n=5$),
and a feasible current of order $10^{-8}A$ is observed,  two order
larger than reported in Refs. \cite{14,20}. Second, the
rectification effect persists in a much wider region than  in
previous investigations \cite{14,20}. The rectification does not
degrade when $n$ is varied.

 The paper is organized as follows. The the device model and computation
method  are presented in Sec. II; the results and discussions are
given in Sec. III; and a conclusion is drawn in Sec. IV.

\section{ Model and Method}

\begin{figure}
\epsfig{figure=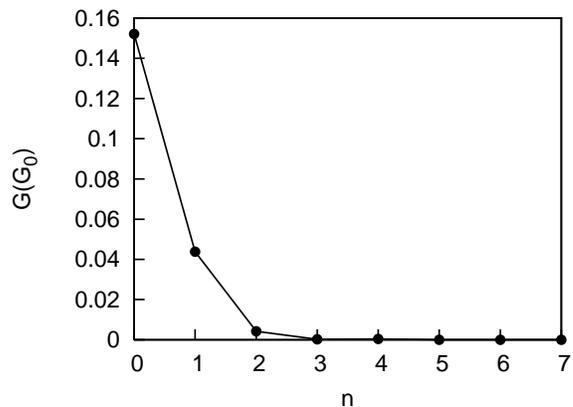,width=3.0in}
 \caption{The equilibrium
conductance as a function of the number $n$.  }
\end{figure}

The model structure for our theoretical analysis is illustrated in
Fig. $1$. The single molecule HOOC-C$_6$H$_4$-(CH$_2$)$_n$ with
sulfur end is sandwiched between two Al(100) electrodes with
finite cross section. The electrodes are chosen from the perfect
Al crystal along the (100) direction, and the number of atoms in
each atom layer is arranged in sequence 5,4,5,4 $\cdots$. Four Al
atomic layers (5, 4, 5, 4) are selected for the electrode cell.
The single molecule HOOC-C$_6$H$_4$-(CH$_2$)$_n$ with sulfur end
together with four surface atom layers in the left electrode and
three surface atom layers in the right electrode is chosen as the
central scattering region, as indicated by two vertical white
lines in Fig. $1$. The structure of every molecule
HOOC-C$_6$H$_4$-(CH$_2$)$_n$ with sulfur end is optimized via
Hartree-Fock approximation in ArgusLab \cite{25,26,27,28}, and the
energy convergence of $10^{-10}$ $kcal$/$mol$ has been achieved.
The molecule-lead distance is fixed to be a constant for all the
systems. The distance of the H-Al(from the left terminal hydrogen
atom to the surface of Al electrodes) is 1.5${\AA}$ and the S-Al
is 1.0${\AA}$.

The calculations for transmission and Current-Voltage(I-V)
characteristics have been performed using a recently developed
first-principles package TranSIESTA-C, which is based on the
nonequilibrium Green's function (NEGF) technique. The TranSIESTA-C,
as implemented in the well tested SIESTA method, is capable of fully
modeling self-consistently the electrical properties of nanoscale
devices, which consist of an atomic scale system coupling with two
semi-infinite electrodes as shown in Fig.1. Such a nanoscale device
is divided into three parts: left and right electrodes, and a
central scattering region. In fact, the central region includes a
portion of the semi-infinite electrodes. The external potential bias
is included in the self-consistent calculation directly. Therefore,
the effects of the bias voltage on the electronic structure of the
system can be fully considered. Details of the method and relevant
references can be found elsewhere \cite{21,22,23,24}. In our actual
calculations, the convergence criterion for the Hamiltonian, charge
density, and band structure energy is $10^{-4}$ and the atomic cores
are described by normconserving persudopotentials.

\begin{figure}
\epsfig{figure=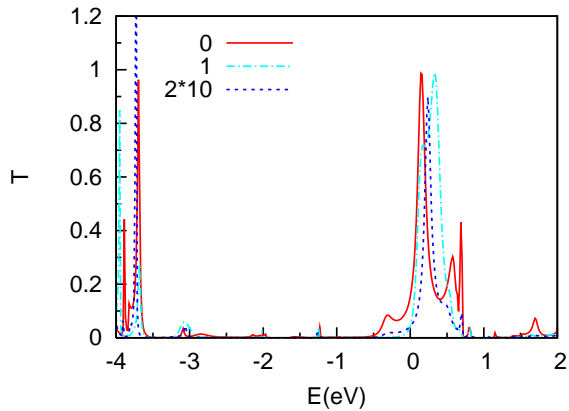,width=3.0in}

\caption{The transmission $T$ as a function of energy E for $n=0$,
1 and 2. For comparison, the transmission for $n=2$ is magnified
by 10.  }
\end{figure}

\section{RESULTS AND DISCUSSIONS}

Firstly, the equilibrium conductance as a function of the number
$n$ of CH$_2$ has been studied and is presented in Fig. $2$. We
find that the conductance decreases exponentially as the number
$n$ of CH$_2$ increases. The conductance is 0.15G$_0$, 0.043G$_0$,
0.0043G$_0$, 0.00026G$_0$ if the number of CH$_2$ $n=0,1,2,3$
(G$_0 =2e^2/h$ is the conductance quanta). The similar exponential
decrease of conductance with the length has been found for carbon
chains\cite{29}.

The $I$-$V$ curves are plotted in Fig. $3$ (the cases $n=4, 6$ are
not shown in this work). As shown in Fig. $3$, the current takes a
large value of order $\mu A$ both at the negative bias and at the
positive bias, When $n=0$ and 1. When $n\ge 2$, interesting
phenomenon are observed. The current through the molecule structure
at the negative bias is increasing slowly irrespective of the  bias
voltage increase, as compared to the quick increase of the current
with the increasing bias voltage in the positive bias case.  When
the bias is positive, the current reaches its maximum value at a
bias, which is different for different $n$ systems. These are
typical current rectification  and NDR phenomena.  A significant
rectification ratio $\sim$38 is found for the molecule system with
the number of CH$_2$ $n=5$ , and the current in the negative bias
side is negligibly small. We note that, for every configuration, a
strong NDR can be observed as the bias is positive. The NDR in our
system may find  potential applications as switches.

\begin{figure}
\epsfig{figure=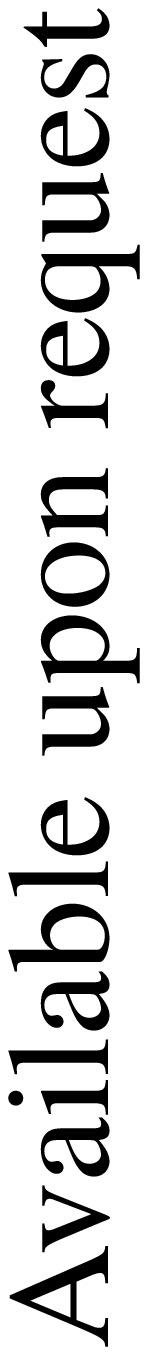,width=3.0in} \caption{ (color online) The
LUMO and HOMO distribution on the molecule for $n=0$,1 and 5
respectively: (a) LUMO for $n=0$, (b)HOMO for $n=0$,(c) LUMO for
$n=1$, (d)HOMO for $n=1$,(e) LUMO for $n=5$ and (f)HOMO for $n=5$.
}
\end{figure}

\begin{figure}
\epsfig{figure=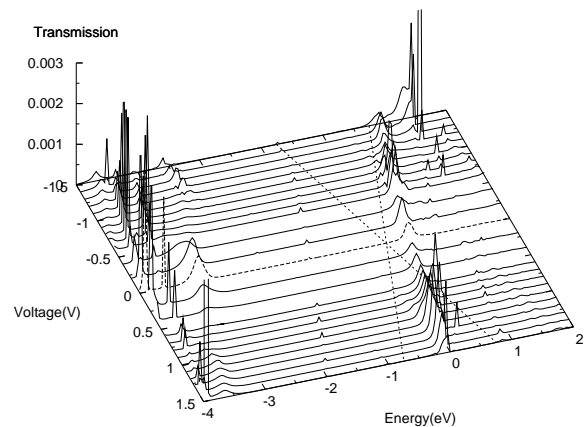,width=3.0in}

\caption{ Transmission spectra, $T(E,V_b)$, for the $n=5$ case.
The two crossing dotted lines indicate the bias window. }
\end{figure}

What are the underlying mechanisms for the current rectification
and NDR in our molecule system? Why the current rectification can
be observed only when the number of CH$_2$  exceeds $1$? To answer
these two questions, one need to know detailed information of the
transmission spectra of such a system. For current rectification
to occur,  two essential conditions should be satisfied that the
transmission resonance must be reasonably sharp and exists only at
one vicinity side of the Fermi energy\cite{30}. The transmission
 spectra $T(E)$ are plotted in Fig. 4 for $n=0$, 1 and 2 cases. We can see
from Fig. 4 that the resonance is rather wide for $n=0$ case and has
a long tail near the Fermi energy. A direct comparison reveals that
the resonance in the case $n=1$ is wider than that in the case
$n=0$, but moves toward right. However, the resonance becomes much
more narrow and can be seen only at the right side of the Fermi
energy for the case $n=2$. The reason is that, when $n=0$, the
sulfur atom is directly attached to the benzene, thus the orbitals
of the lone electron pair of the sulfur are overlapped with that of
the $\pi$ electrons of the ring, resulting in molecular orbitals
distributed almost over the ring and the sulfur. Such an observation
is confirmed in Fig. $5(a)$, where both the LUMO and HOMO extend
almost over the whole molecule. Since the sulfur is directly coupled
to the electrode, the molecular orbitals are broadened. In the $n=1$
case, the sulfur is separated from the ring by a single insulating
-CH$_2$- group. However, the separation is not large enough to
prevent elongated $p$ orbitals of sulfur and $sp_3$ hybridization of
carbon from direct overlapping, as seen in Fig.$5(b)$ where the LUMO
and HOMO still distribute almost over the whole molecule. As a
result, the electronic level is still much broadened. Things changes
when $n=2$. Two insulating groups move the sulfur away from the ring
by 4.09{\AA}, so that the direct overlap between the sulfur and the
ring wave functions is weakened, and the LUMO and HOMO do not
distribute on the left (CH$_2$)$_2$ (which is not shown). That's why
rectification can be observed only for $n\ge 2$.

To probe clearly the origin of rectification and NDR, we take the
molecule HOOC-C$_6$H$_4$-(CH$_2$)$_5$ as an example. When the
molecule is coupled two Al electrodes, the calculation shows that
the LUMO and Homo energies are 0.394$eV$, $-$2.896$eV$,
respectively, consistent with the indications in the transmission
spectra in equilibrium in Fig. $6$.  We can see from Fig.5$(c)$
that the LUMO distributes over the HOOC and the carbon atoms in
benzene ring, and the HOMO mainly on the sulfur atom. The electron
has an accessible LUMO near the Fermi energy only of one
electrode,  the left electrode in the present work. This asymmetry
leads to a significant current when the applied bias is positive
and a negligibly small current in the negative case.

The calculated transmission spectra $T(E,V_b)$ are presented in
Fig.$6$ for different applied bias voltages from $-1.5V$ to
$+1.5V$ with an interval $0.1V$ and a doubled interval within
[-0.6,0.6]V. The dashed line represents the transmission spectrum
in equilibrium. The current is obtained using the
Landauer-B$\ddot{u}$ttiker formula:
$I=\int_{{\mu}_L}^{{\mu}_R}T(E,V_{b})dE$. The bias window
contributing to the current integral is indicated by two dotted
lines in Fig.$6$. The transmission resonance in equilibrium
appears on the right side of the Fermi energy, and its width is
narrow, about 0.13 $eV$. With the increase of the positive bias,
the resonance moves toward the bias window and makes contributions
to the current. As a result, the current increases rapidly and
reaches its maximum value at the bias 1.2 $V$. However, when the
applied bias is negative, the resonance moves far away from the
Fermi energy as the bias increases, and finally is pushed out of
the bias window. It  results in a negligibly small current. This
is a typical phenomenon of current rectification. A notable
rectification ratio of ~38 can be observed at the bias 1.2 $V$.
When the positive bias continues to increase, the transmission
within the bias window is degraded in magnitude and width, leading
to an NDR phenomenon.

\section{CONCLUSION}

A system of HOOC-C$_6$H$_4$-(CH$_2$)$_n$  is studied systematically
using a first principles method based on NEGF. The equilibrium
conductance decreases exponentially with the increase of $n$. Strong
rectification effect is observed when $n\ge2$ and a significant
rectification of $\sim$38 is achieved from $n=5$ for bias $<$1.2$V$.
Irrespective of the number of CH$_2$, a strong NDR can be seen under
positive bias.  A detailed analysis on the origin of rectification
is given from the distribution of the HOMO and LUMO and from the
transmission spectra at various bias.

This work was supported by the NSFC under Grant No. 10404010, the
Project-sponsored by SRF for ROCS, SEM, and Scientific Research
Fund of Jiangxi Provincial Educational department(112[2006]), and
the Talent Fund of Jiangxi Normal University.

\end{document}